\documentclass[runningheads]{cl2emult}

\usepackage{graphicx} 

\begin{document}
\title*{Planar Integrated Optics and 
astronomical interferometry\\ 
{\small Optique Int\'egr\'ee planaire pour l'interf\'erom\'etrie 
appliqu\'ee \`a l'astronomie}}

\titlerunning{Planar Integrated Optics and 
astronomical interferometry}

\author{Pierre Kern \inst{1}
\and Jean Philippe Berger \inst{2}
\and Pierre Haguenauer \inst{2} \inst{3}
\and Fabien Malbet \inst{1}
\and Karine Perraut\inst{1}}

\authorrunning{Pierre Kern et al.}
\institute{Laboratoire d'Astrophysique de l'Observatoire de Grenoble 
BP 53, 38041 Grenoble Cedex 9, France
\and Laboratoire d'Electromagn\'etisme Microondes et Opto\'electronique 
BP 257, 38016 Grenoble Cedex 1, France
\and CSO mesure,70 rue des Martyrs, 38000 Grenoble, France}
\maketitle

\begin{abstract}  
  Integrated optics (IO) is an optical technology that allows to 
  reproduce  optical circuits on a planar substrate. Since 1996, we have investigated
  the potentiality of IO in the framework of astronomical single mode
  interferometry. We review in this paper the principles of IO, the
  requirements for interferometry and the corresponding
  solutions offered by IO, the
  results of component characterization and the possible fields of
  application.

\end{abstract}

\begin{abstract}  
  L'optique int\'egr\'ee est une technologie qui permet de reproduire des
  circuits optiques sur un substrat planaire. Depuis 1996, nous menons des
  recherches sur les potentialit\'es de l'optique int\'egr\'ee dans le
  contexte de l'interf\'erom\'etrie monomode en astronomie. Dans cet
  article, nous passons en revue les principes de l'optique int\'egr\'ee,
  les sp\'ecifications propres \`a l'interf\'erom\'etrie et les 
  solutions correspondantes
  offertes par cette technologie, les r\'esultats de caract\'erisations de
  composants ainsi que les domaines d'application.

{\bf Mots cl\'es :} interf\'erom\'etrie, synth\`ese d'ouverture optique,
optique int\'egr\'ee, optique planaire, optique monomode.
\end{abstract}
\section{Introduction}

The use of guided optics for stellar interferometry was introduced to
reduce constraints while combining coherent beams coming from several
telescopes of an interferometer. Claude Froehly proposed in 1981
\cite{Froe} to use single mode fibers to solve the problems linked to the
beam transportation and high number of degrees of freedom of such an
instrument.  Laboratory developments \cite{Sha1} \cite{Rey1} and on the sky
experiments \cite{Cou1} have shown the important improvements introduced by
guided optics.  More than experimental setup simplification, single mode
guided optics introduces modal filtering which allows translation of the
phase disturbance for the incoming wavefronts into calibrable intensity
fluctuations.

The analysis of existing fiber-based experiments led us to propose planar
integrated optics (hereafter IO) as a solution for some of the remaining
difficulties linked to fiber optics properties or to the general instrument
setup \cite{Ker1},\cite{Ker2}.  In this paper, we present a review of the
work done since 1996 by the team composed of partners from research
laboratories (Laboratoire d'Astrophysique de l'Observatoire de Grenoble -
LAOG, CEA/LETI and Laboratoire d'Electromagn\'etisme Microondes et
Opto\'electronique - LEMO) and from industrial laboratories (GeeO, CSO).
Section 2 presents the principle of planar IO and an introduction to the
related technology. This section presents also typical available IO
functions. Section 3 summarizes the requirements for an interferometric
instrument and the solutions offered by IO. Section 4 presents the results
obtained during that period and shows how IO can be a convenient solution
for the instrumentation for interferometry.  The last section gives some
perspectives of our developments.

\section {Integrated Optics technology}
\subsection{ Planar optics principle}

Optical communications using single mode fiber optics for long
distance connection impose periodic signal amplification. Triggered
by telecommunications industry requirements, major efforts have been
done to manufacture compact single chip repeaters directly connected
to fibers avoiding multiple signal conversion \cite{Mill}.This has
lead to develop techniques able to integrate complex optical circuits,
as for integrated electronics, on small chips. The main technological
breakthrough resided in the ability of integrating single-mode
waveguides in a given substrate.

In a planar waveguide (Fig.\ 1a) optical guidance is guaranteed by the
three step-index infinite planar layers ($n_2 > n_1$ and $n_3$)
\cite{Jeun}.  The core layer thickness of index $n_3$ ranges between
$\lambda/2$ and $10 \lambda$ depending on index difference $\Delta n$.
A large $\Delta n$ leads to better light confinement.
\begin{figure}[t]
  \centering
  \includegraphics[width=1\textwidth]{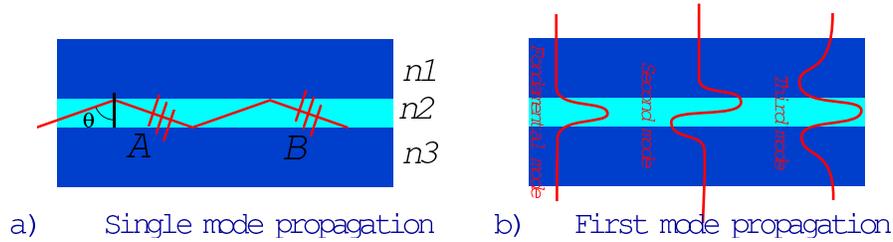}
  \caption[]{Principle of mode propagation within a waveguide.}
  \label{SM-WG}
\end{figure}
A full electromagnetic field description, shows that the modal beam
propagation applies in waveguide structure \cite{Jeun}. The main part of
the carried energy lies in the waveguide core, but evanescent field
propagates in lateral layers and contribute to the mode propagation.  A
guiding structure with a given thickness and layers refractive index is
characterized by a cut-off wavelength $\lambda_{c}¥$, separating the single
mode propagation ($\lambda > \lambda_{c}¥$) where only the fundamental mode
propagates and the multimode propagation condition ($\lambda \leq 
\lambda_{c}$, Fig.\ 1b).

Only the single mode regime is considered in our developments. However
multimode guided structures have been tested \cite{Sha2} for stellar
interferometry.

\subsection{Waveguide manufacturing}

The guided area is obtained by {\it ion exchange technique \cite{Laur}}.
The Na+ ions of the glass substrate are exchanged by diffusion process with
ions K+, Ti+, Ag+ of molten salts and result in an increase of the
refractive index, producing the three-layer structure (air / ions / glass)
capable to confine vertically the light.  The implementation of the optical
circuit is obtained by standard photomasking techniques (see Fig.\ 2 left)
to ensure the horizontal confinement of the light. While ion exchange
occurs at the surface of the glass, an additional step of the process can
embed the guide, either by applying an electric field to force the ions to
migrate inside the structure or by depositing a silica layer. The waveguide
core is the ion exchange area and the cladding the glass substrate or the
glass substrate and air. Depending on the type of ions, $\Delta n$ can
range between 0.009 and 0.1. This technology produced in Grenoble by LEMO
and GeeO/Teem Photonics is commonly used for various components used in
telecom and metrology applications.

The waveguide structure can also be obtained by the {\it etching} of silica
layers \cite{Mott} of various refracting indices (phosphorus-doped silica
or silicon-nitride). As for other techniques photomasking is required to
implement the optical circuit (see Fig.\ 2 right). The manufacturing
process allows to choose either a high $\Delta n$ ($\Delta n \geq 0.5$) to
implement the whole circuit on a very small chip with small radii curves,
or very low ($0.003 \leq \Delta n \leq 0.015$) for a high coupling
efficiency with optical fibers. This technology is used at CEA / LETI to
produce components for various industrial applications (telecommunication,
gyroscopes, Fabry-P\'erot cavities or interferometric displacement
sensors).

Single mode waveguide structure are also produced by UV light
inscription onto {\it polymers}.  The transmission of the obtained
components are still too small for our applications. A technology
based on $LiNbO_{3}$ cristal doping by metals allows to produce
single-mode waveguides with interesting electro-optical properties but
this has not been tested yet for our specific applications.
\begin{figure}[t]
\centering
\includegraphics[width=\textwidth]{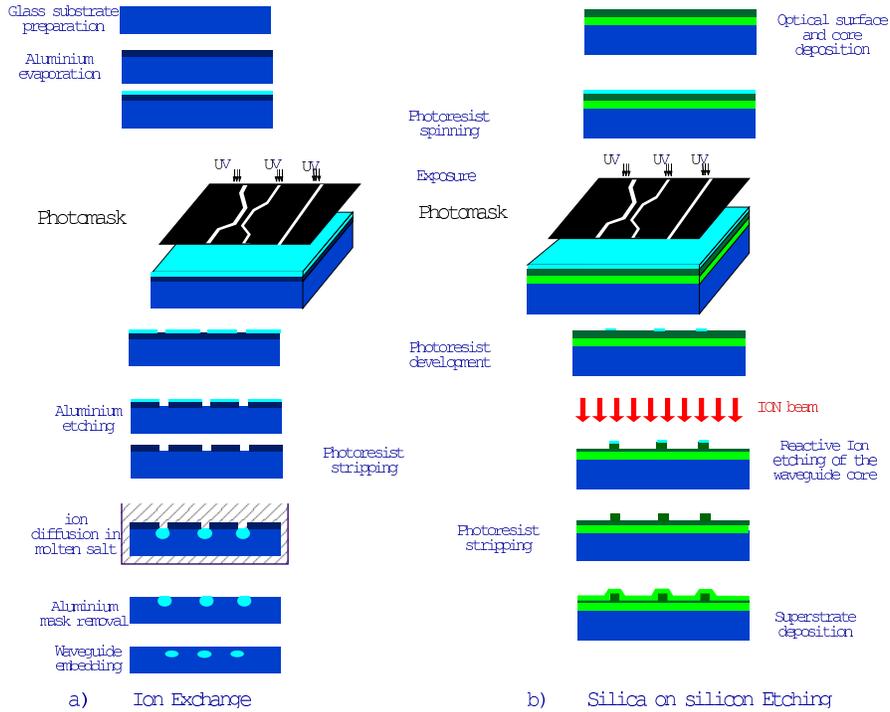}
\caption[]{Ion exchange (LEMO) and etching (LETI/CEA) techniques}
\label{Ion-ex}
\end{figure}

\subsection{Planar optics functions}

Several functions working at standard telecom wavelengths ($0.8\mbox{
  $\mu$m}$, $1.3\mbox{ $\mu$m}$ and $1.5\mbox{ $\mu$m}$) are available
  with the different technologies.  Figure 3 displays an example of IO
  chip used in an interferometric displacement sensor \cite{Lang}. It
  nicely illustrates IO capability because it contains nearly all 
  basic functions. The reference channel of the interferometer head
  and the measurement channel are provided by splitting the He-Ne
  light by a direct Y-junction. The light of the measurement channel,
  retroreflected is injected again in the waveguide and directed to
  the interferometer head thanks to a directional coupler. The
  interferometer head is a large planar guide fed by two tapers.
  Interference between the two beams produces fringes which are
  sampled by four straight guides, providing measurements with
  $\lambda {\rm /4}$ phase difference at the same time. The most
  common functions are listed below:
\begin{itemize}
    \item Direct Y-junctions for achromatic 50/50 power splitting.
    \item Reverse Y-junctions for elementary beam combination as bulk
    optics beam splitters with only one output. The flux to the second
    interferometric state in phase opposition is radiated into the
    substrate.  
        \item Directional couplers allow the transfer of the
    propagated modes between neighbour guides. The power ratio division
    in each output beam is linked to the guide
    separation, the interaction length and the wavelength. Symmetrical
    couplers ensure a chromatic separation of the signal. Achromatic
    separation requires an asymmetrical design.  
\item X-crossings
    with large angles ($\geq 10^{\circ}$ ) for guide crossing with
    negligible cross-talk effects. Smaller angles favor power exchange
    between the guides; 
\item Straight waveguides.  
\item Curved
    waveguides give flexibility to reduce the component size.
    Possible curvature radii depend on the core and substrate index
    difference.  
\item Tapers or adiabatic transitions, thanks to smooth transition of
the guide section, adapt propagation from a single mode straight
waveguide to a larger waveguide. Consequently light propagates and
remain in the fundamental mode of the multimode output
waveguide. These components reduce the divergence of the output beam.
\end{itemize}
\begin{figure}[t]
\centering
\includegraphics[width=\textwidth]{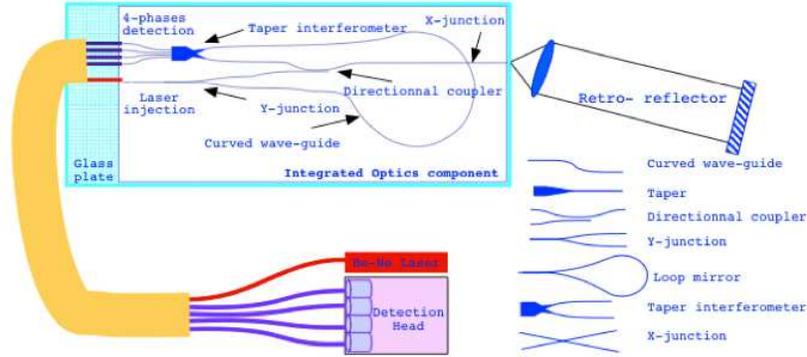}
\caption[]{Planar Optics displacement sensor developed by LEMO 
  \cite{Lang} and corresponding function list.}
\label{displ-sens}
\end{figure}

\section{Instrumental requirements in stellar interferometry}

\subsection{ Functional requirements}

The wavefront distortion, due to atmospheric transmission or to
instrumental aberrations, induces fringe visibility losses. The main
errors can be corrected by an adaptive optics system or by an
appropriate optical design adjusting the entrance pupil diameter to
the local value of atmospheric coherence area diameter for the
considered wavelength. The remaining phase errors on the incoming
wavefront can be removed using a {\it spatial or modal filtering}
\cite{Cou2}. The high spatial frequencies introduced by the wavefront
distortion are rejected by a field stop in the Fourier plane.  The
diffraction limited image of an unresolved object leads to a beam
\'etendue of $S \Omega = \lambda^{2}¥$. It corresponds to single
mode propagation with no longer pupil and focal plane position. This
fundamental mode is properly directed by a single-mode waveguide
(fiber optics or planar waveguide) when it meet the $\lambda^{2}¥$
beam \'etendue condition.  If higher modes enter the waveguides they are
rejected out of the core according to propagation laws. This modal
filtering is wavelength dependent.  Current simulations are in
progress \cite{Mege} to determine the appropriate dimensions of the
guide to operate an efficient filtering: waveguide diameter, required
guide length, operating wavelength range.

When a {\it modal filtering } is applied, phase distortions are translated
into intensity variations in the waveguide. To calibrate the contrast of
the fringe pattern, a photometric correction can be applied on the recorded
interferometric signal taking into account the flux variations for each
incoming beam. The variation of the telescope fluxes are monitored together
with the interferometric signal. Associated to the modal filtering, {\it
  photometric calibration} allows significant improvements of the fringe
visibility estimation. This principle has been applied successfully with
accuracy down to 0.3\% with the FLUOR instrument \cite{Cou1}.

Observation over {\it atmospheric spectral bands} are required
in many applications. Wavelength dependent parameters may affect the
extraction of the interferometric signal. The bias introduced by
differential chromatic dispersion between interferometer arms due to optical
components must be compensated or calibrated.

Instrumental differential rotations and phase shifts between the
polarization directions can affect the fringe visibility. Symmetric optical
design allows to reduce polarization effects but high contrasts require the
compensation of the {\it differential effects on polarizations \cite{Rou1}}
by Babinet compensators suggested by Reynaud \cite{Rey2} 
or Lef\`evre fiber loops.

The {\it optical path equalization} is necessary from the interference
location to the stellar object with a sub-micrometer accuracy. Delay lines
operates this optical function. Fiber optics solution have been proposed
and tested in laboratory \cite{Simo}, \cite{Zhao}.  Differential fiber
dispersion remains the limiting factor of the proposed concept.  ESO
prototype fringe sensor unit \cite{Rabb} uses a fiber optics delay line.

Finally, {\it very high OPD stability} is mandatory, especially for phase
closure.  Variations of the OPD leads to phase relationship loss and reduce
the image reconstruction capability. The opto-mechanical {\it stability} of
the instrument strongly affects the fringe complex visibility.

\subsection{System requirements}
The {\it beam combiner} ensures visibility and phase coding of the
interference pattern. Telescope combiner for more than 3 telescopes are
required to obtain synthesized images. The image reconstruction implies
very accurate phase difference control between the interferometer arms.
The beam combination can be done either using single mode or multimode
optical field. In each case, the combination is performed using coaxial or
multi-axial beams.

The {\it spectral dispersion} of the fringes is used for either
astrophysical parameter extraction, or for fringe detection.  Stellar
interferometry is generally performed within the standard atmospheric
spectral windows of ground-based observation.  The spectral analysis is
achieved either by using optical path difference modulation (double Fourier
Transform mode) \cite{Mari} in coaxial mode or with dispersive components.
In the latter case, the fringe light is focused on the spectrograph slit
using a cylindrical optics \cite{amber} to concentrate the flux along the slit.

A {\it fringe tracker} allow longer acquisition times and increase the
instrument sensitivity.  Time dependent behaviors affect the central white
fringe position: sidereal motion, instrument flexures and fine telescope
pointing decay on smaller scale. Finally atmospheric turbulence affects
ground based observations inducing atmospheric piston at the interferometer
baseline scale. The fringe tracker ensures the fringe stability thanks to
a suitable delay-line controlled with a proper sampling of the OPD
fluctuations at a frequency compatible with the considered time scale.  It
avoids visibility losses due to fringe blurring.  The fringe sensor
is part of the fringe tracker, it is
aimed to measure the central fringe location of the interference pattern
with suitable accuracy. Various principles have been proposed \cite{Shao},
\cite{Koeh}, \cite{Cass}.  Multi-axial mode allows a complete sampling in a
single acquisition, while coaxial mode requires an OPD active modulation.

Astrometrical mode requires milli-arcsecond positioning accuracy on
simultaneous observations on two distant stars. Such measurements require
an appropriate {\it metrology control} of the optical path length for the
two stars, from the telescope entrance to the fringe detection device.

More recently for search of faint objects around bright stars, 
instruments using
interferometry have been proposed \cite{Brac}. The on-axis star light is
extinguished thanks to a $\pi$ phase delay on one of the
interferometer  arms
before combination, providing a {\it nulling interferometer}. Fringe
separation is adjusted to place the central fringe of the off-axis searched
object interference pattern on the black fringe position. Interferometer
pupil arrangement is optimized to obtain enhanced central star light
rejection.

\subsection{IO: a promising solution}

Intrinsic properties of planar IO solve a large part of the functional
requirements described above of an instrument dedicated to interferometry
mostly due to its ability to propagate only the fundamental mode of the
electromagnetic field:
\begin{itemize}
\item Single mode propagation within a half octave without significant
  losses \cite{Hag1}.
\item Broad band transmission if the used functions are compatible with an
  achromatic transmission (available for individual J, H and K atmospheric
  bands) \cite{Hag1}, \cite{Sev1}.
\item Intrinsic polarization maintaining behavior for most of the cases
  \cite{Ber2}.
\item High optical stability on a single chip \cite{Ber2}.
\item Accurate optical path equality ($leq 2 \mu{m}$) if suitable care is applied in the
  design and in the component manufacturing \cite{Hag2}.

\item Reduced differential effects between interferometric paths, since
  they are all manufactured during the same process on the same substrate
  \cite{Hag2}.
\item Photometric calibration easily implemented using existing basic
  functions (direct Y junction, directional coupler)
\item Measurements performed at LEMO shows that temperature constraints
  applied on a component only introduces $\lambda/90000 \;/\mbox{
    mm}/^{\circ}C)$ phase shift compatible with phase closure requirements
  \cite{Ber2}.
\end{itemize}
   
Moreover, existing IO systems are also able to provide part of the required
subsystem for interferometry in astronomy:
\begin{itemize}
\item Beam combination can be achieved either in coaxial mode by reverse
  Y-junction, directional couplers, Multi Mode Interferometer
  \cite{Elsa} (MMI) and/or by tapers in multi-axial mode (see Fig.\ 4).
\item The output of a taper interferometer produces an illumination along a
  single direction suitable for spectrograph entrance slit. 
\item Using beam combiner for fringe sensing, many solutions may be applied:
  \begin{itemize}
  \item A coaxial combiner with 2 outputs in phase opposition, or 4 outputs
    with a $\pi /4$ phase difference provides the proper fringe
    sampling \cite{Sev3}. More phase samples can be provided by small OPD
    modulation.
  \item A suitable sampling of the dispersed output of a taper
    interferometer can also be used on an array detector for a fringe
    sensing using the Koechlin arrangement \cite{Koeh}.  
  \end{itemize}
\item Several interferometric displacement sensors have been proposed and
  even offer for sell using planar optics. The existing concepts are
  directly applicable to metrology for interferometry \cite{Lang}.  
\item Nulling interferometry requires a $\pi$ phase shift. Interferometer
  including such functionality has been tested for industrial use with
  laser. 
\end{itemize}
\begin{figure}[t]
\centering
\includegraphics[width=.8\textwidth]{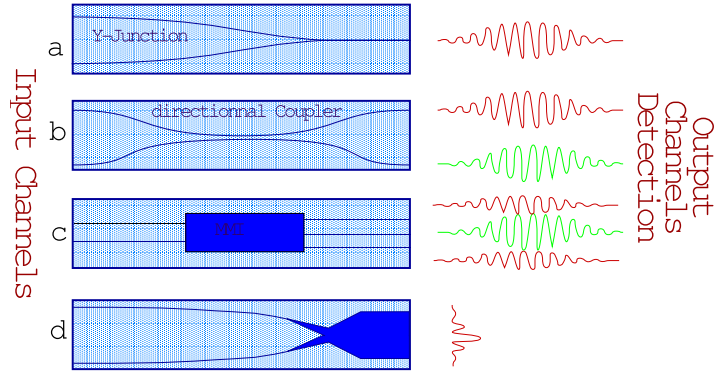}
\caption[]{Available beam combiners. For the 
reverse Y-junction (a) part of the signal, in phase opposition, is radiated inside the 
substrate.
Directional coupler (b) provides two outputs in phase 
opposition.
MMI design (c) corrects this functional loss, providing three
outputs. Each external guide collects 1/4 of the interferometric
signal and provide modulated light in phase opposition with the
central guide (which contains 1/2 of the interferometric signal). The
taper interferometer (d) ensures spatial encoding of the fringes whose sampling
depend on the two taper characteristics (angles,
dimensions).}
\label{beam-comb}
\end{figure}

\section {Main results}

A strong collaboration between LAOG and IO specialists (LEMO, LETI, GeeO,
Teem Photonics) have given us the oportunity to propose a complete development
program.  Starting from available off-the-shelves components, we
demonstrated the ability of the concept to meet the requirements of
interferometry \cite{Malb}, \cite{Ber1}. From the first analysis, we
have developed adapted components for astronomy, with suitable designs.
Systematic tests were performed on first set of components realized with
ion exchanged and etching technique \cite{Hag1}, \cite{Sev1}. We are
starting now developments leading to new functions or new waveguide
technology.  We will then be able to propose first concepts of fully
integrated instrument.

Theoretical investigations are also in progress to optimize the 
combiner parameters \cite{Mege} and analyze the influence of guided beam 
on interferometric scientific data.

\subsection{Instrumental testbeds}
Systematic measurements have been performed on our components to check
their ability to fulfill interferometry requirements:

\begin{itemize}
    \item Photometric measurements to characterize all our components
    transmission over the whole spectral band and the transmission 
    of each function implemented on the corresponding substrate.
  \item A waveguide mode characterization and cut-off wavelength
    determination has been performed thanks to
    an analysis of the intensity spread out in the image of the exiting mode
    and  spectral analysis of the output signal for the whole
    bandwidth \cite{Hag1}.
    \item A Mach Zehnder bench is used for interferometric qualification 
    (Fig.\ 5a). A collimator illuminated by a fiber optics produces a 
    plane wave. It is illuminated either with a He-Ne laser (alignment 
    needs), a laser diode (fringe localization) or a white light source 
    (broad band characterization). The collimated light is splitted by one or 
    several bulk optics beam splitters. The provided channels 
    are imaged on fiber connected to the inputs of the component. 
    \item A stellar interferometer simulator (Fig.\ 5b) for phase closure 
    characterization. In this bench the 
    incoming wavefront is sampled by several apertures with baseline / 
    pupil ratio compatible with real stellar interferometer conditions. 
    It provides unresolved images of a complex object for individual 
    aperture, which can be resolved by the simulated baselines.
  \item In any case the polarization behavior has to be carefully
    characterized. For all of these test benches, polarizers associated
    with polarization maintaining fibers ensure the polarization control of
    the light.
\end{itemize}
\begin{figure}[t]
\centering
\includegraphics[width=1\textwidth]{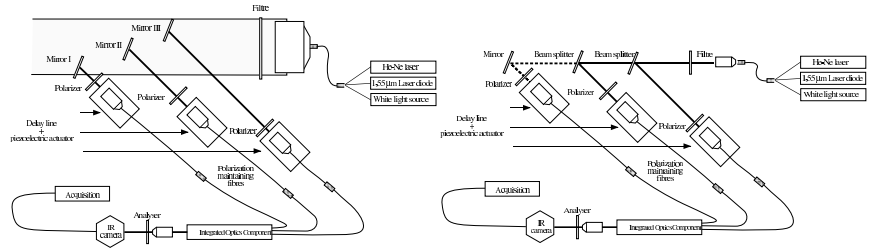}
\caption[]{Interferometric characterization benches:
a) Mach Zehnder bench (amplitude separation ) for interferometric 
characterization of the components (right)
 b) Interferometer simulator (wavefront separation) for image 
reconstruction (left).}
\label{Car-ben}
\end{figure}
       
\subsection{Results with off-the-shelves components}

\begin{figure}[t]
\centering
\includegraphics[width=0.3\hsize]{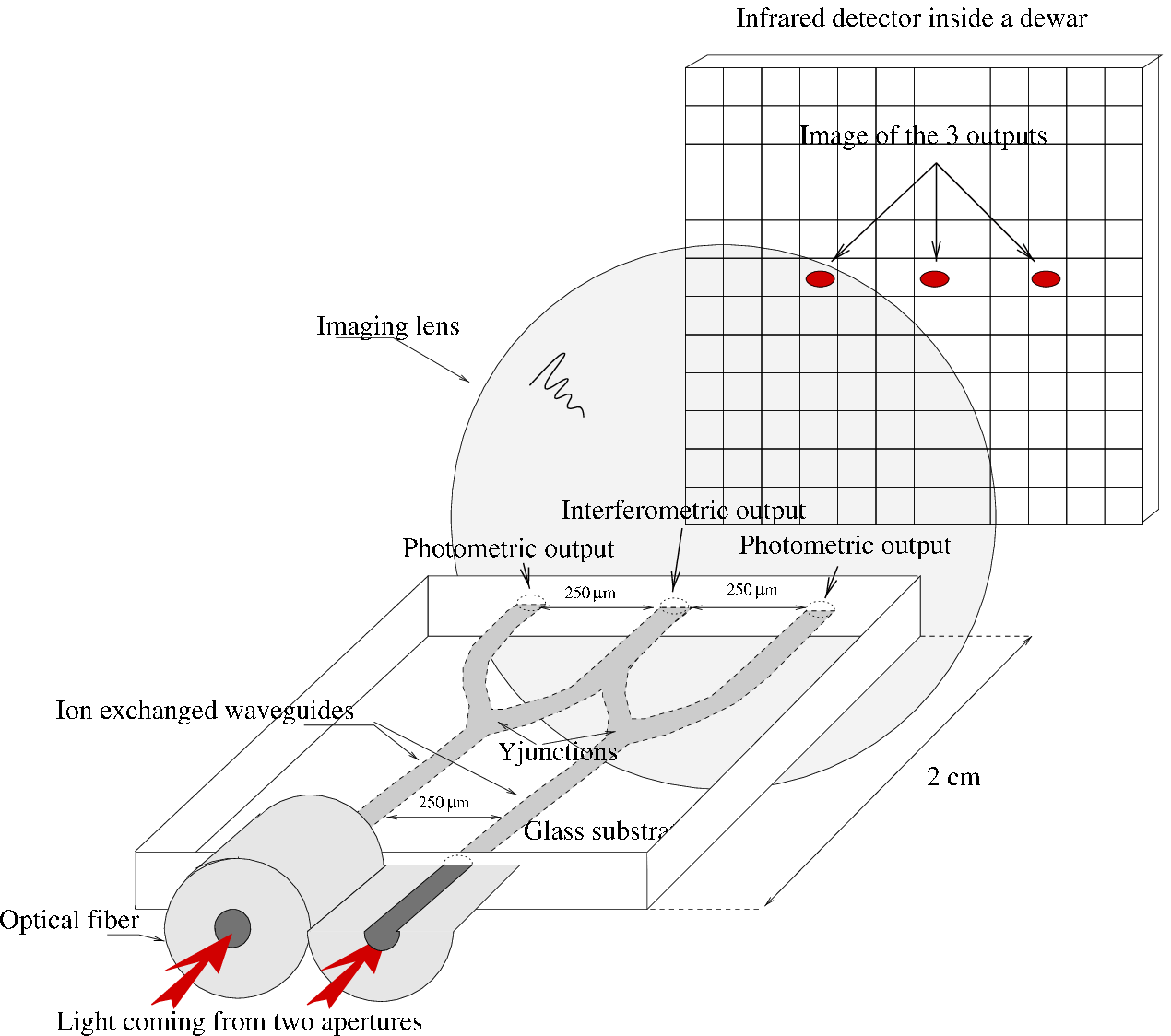} \hfill
\includegraphics[width=0.3\hsize]{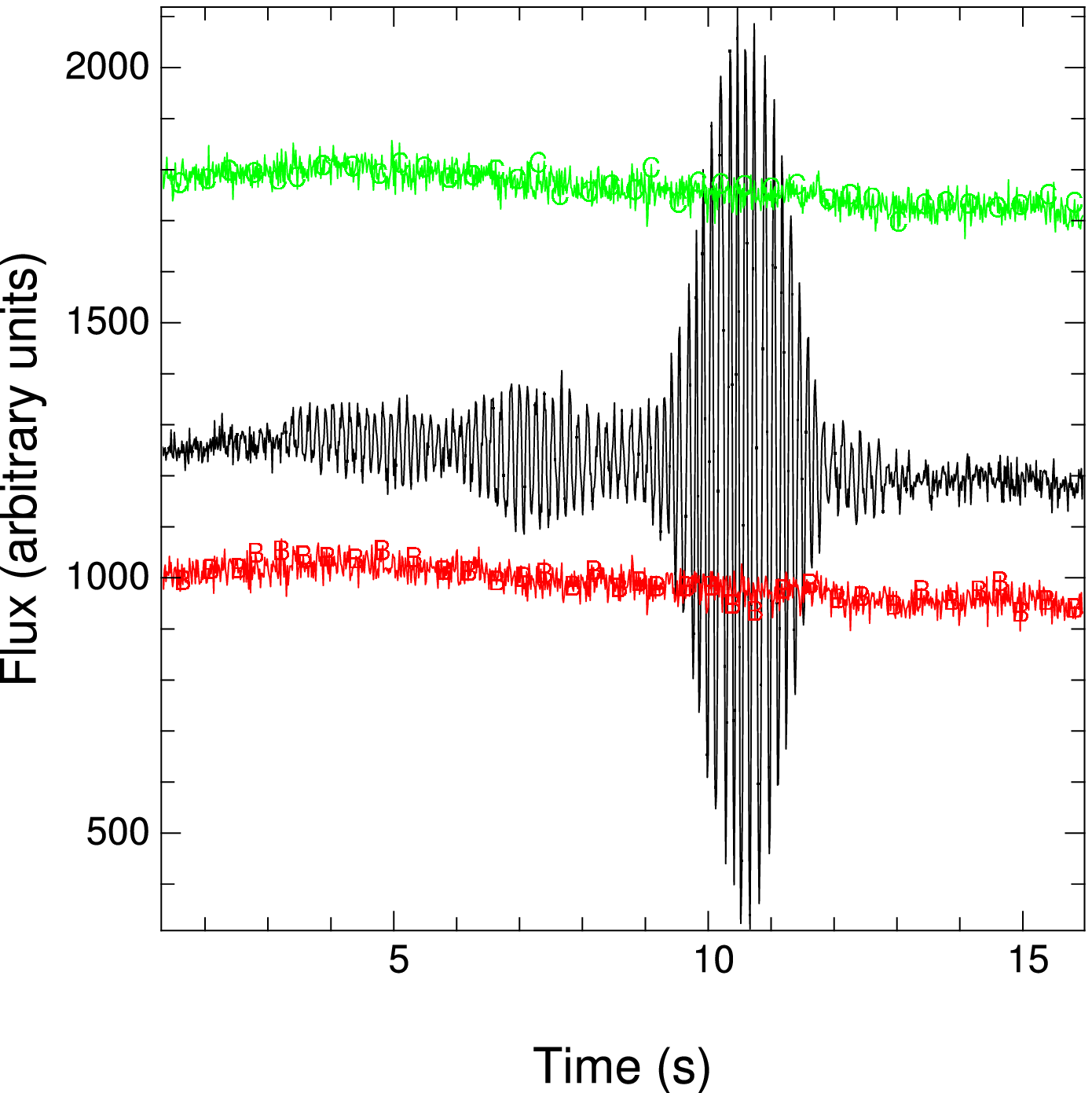} \hfill
\includegraphics[width=0.3\hsize]{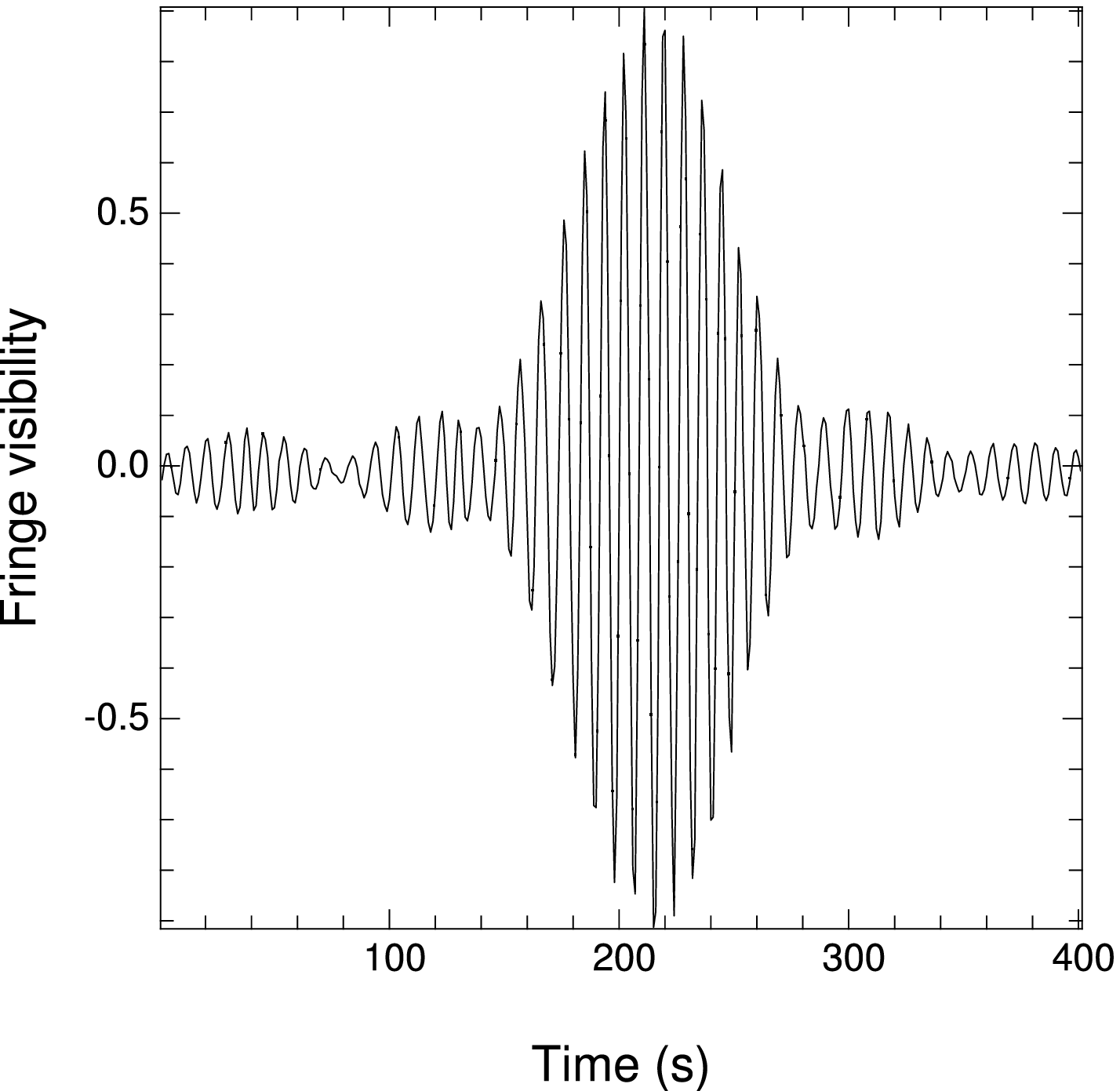}
\caption[]{Coaxial beam combiner (LEMO/GeeO component) and 
corresponding broad band interferogram before correction with both photometric 
channels (left) and corrected from photometric fluctuations (right) \cite{Ber1}, \cite{Hag1}.}
\label{LEMO-comp}
\end{figure}

For a first validation step, part of an existing component was used in
order to combine two beams in coaxial mode. This component contains 2
direct Y-junctions providing a 50/50 beam splitting for photometric
calibration for both inputs and one reverse Y-junction to provide the
interferometric combination.  This component obtained by ion exchange
technique (Ag +) was fully characterized \cite{Ber1}, \cite{Hag1} (see
Figure \ref{LEMO-comp}).

For these tests special care were given to the components / fiber optics
connection. These components were produced and connected by GeeO.  It
provides 92\% fringe contrast on the whole H atmospheric band. The
optimized contrasts was obtained thanks to excellent polarization behavior,
with a contrast stability over several hours as low as 2\%.  The
photometric transmission is 43\%. The main loss of this component is due to
the Y-junction. Using optimized component, this transmission can reach 60\%
with Y-junction and 80\% transmission with an adapted beam combiner
\cite{Hag1}, \cite{Sev2}, \cite{Ber1}.

\subsection{Design of specific components}
Based on those encouraging results we designed specific masks 
for broad band and low flux level uses. 
Components for 2, 3 and 4 telescope combination were produced with 
both silica-on-silicon and ion exchange techniques.
The purpose of this work is to provide a realistic comparison of several 
beam combiners possible designs \cite{Hag2}. 
It is the first time an experimental program can cover such a wide
range of multi-axial and coaxial solutions, pair-wise and all-in-one 
solutions.
 
The component geometry is chosen to match the fiber and detector 
dimensions.

\begin{figure}[t]
\centering
\parbox[c]{0.3\hsize}{\includegraphics[width=\hsize]{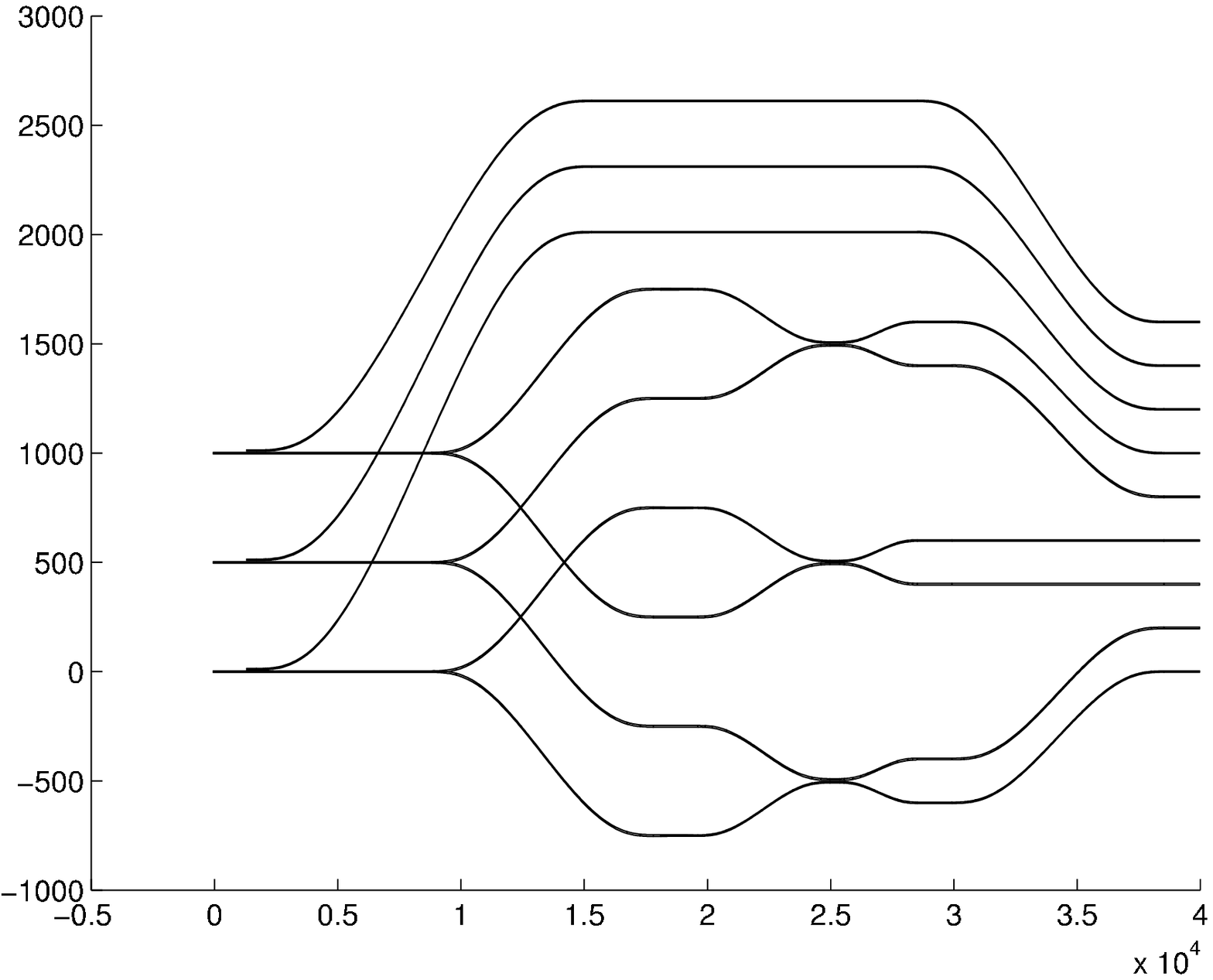}}\hfill
\parbox[c]{0.3\hsize}{\includegraphics[width=\hsize]{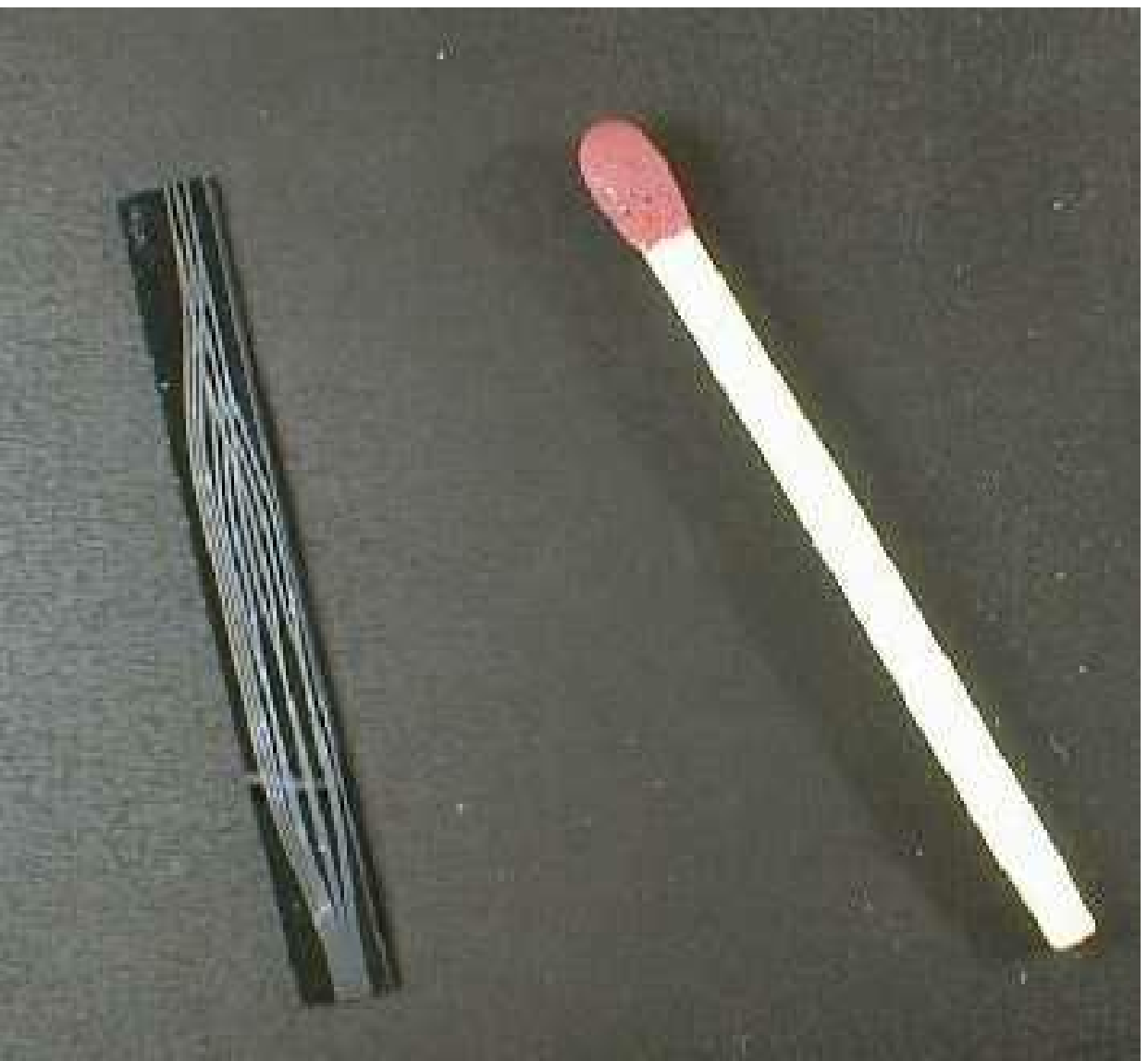}}\hfill
\parbox[c]{0.3\hsize}{\includegraphics[width=\hsize]{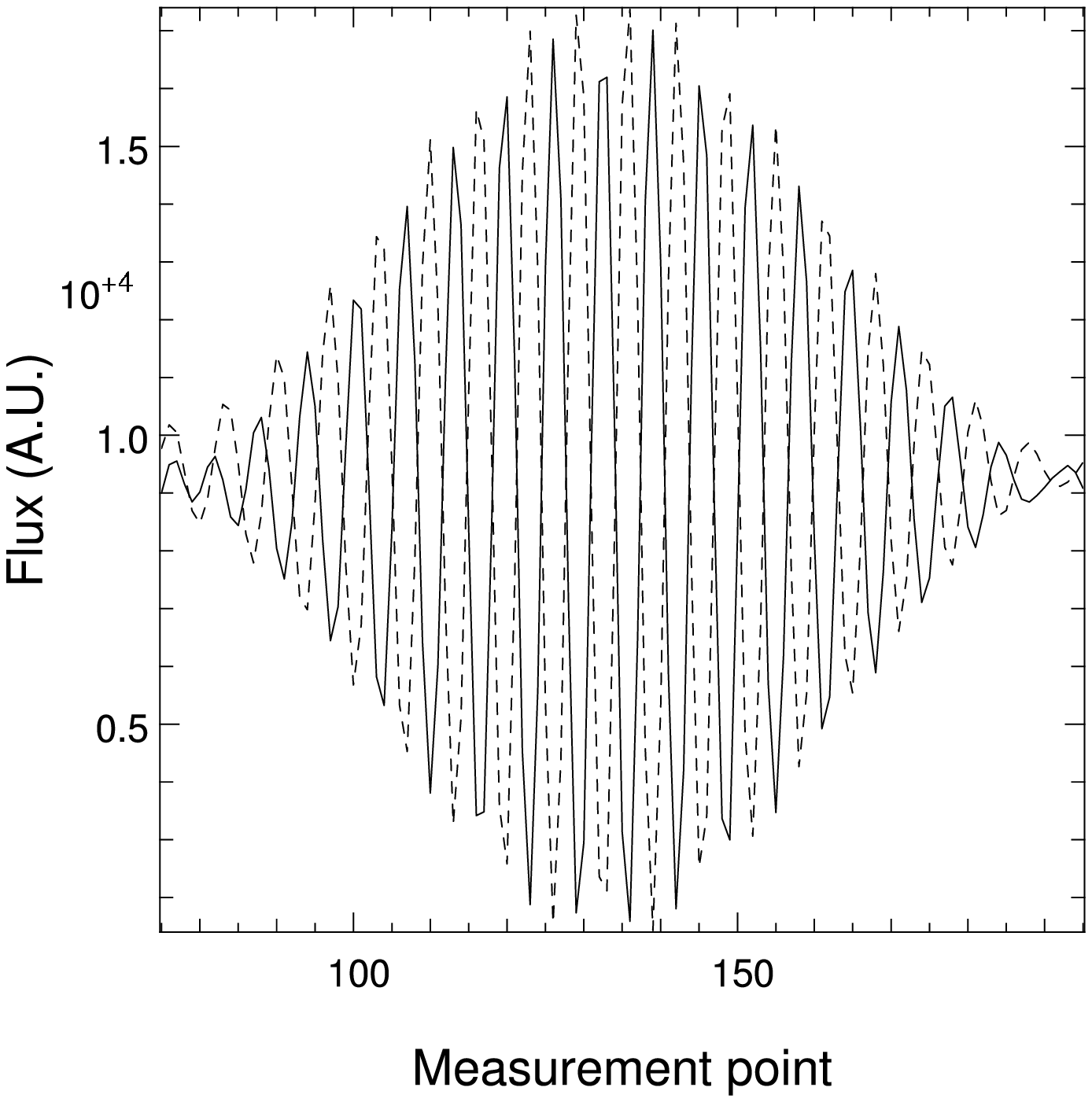}}
\caption[]{3 telescope beam combiner produced using LETI facilities. 
Beam combiners are asymmetrical couplers optimized for the 
whole H band \cite{Sev1}. The obtained interferograms show 90\% 
contrast through the H band, and a throughput higher than 60\%}
\label{LETI-comp}
\end{figure}

The LETI components (silica-on-silicon) allowed the characterization
of asymmetrical couplers, by a systematic analysis couplers parameters
influence.  These components provide two interferometric outputs in
phase opposition.  The obtained design allows quasi-achromatic
splitting ratio
for the whole H band \cite{Sev1}. Component limitations were
identified in terms of transmission and chromaticity and will be taken
into account for next realizations \cite{Sev2}. One of the obtained
components produce pair-wise combinations of 3 telescopes input beams
(see Figure \ref{LETI-comp}). Its excellent performances lead us to propose it
as a solution for the IOTA 3-way new beam combiner.

\begin{figure}[t]
\centering
\includegraphics[width=.6\textwidth, height=2cm]{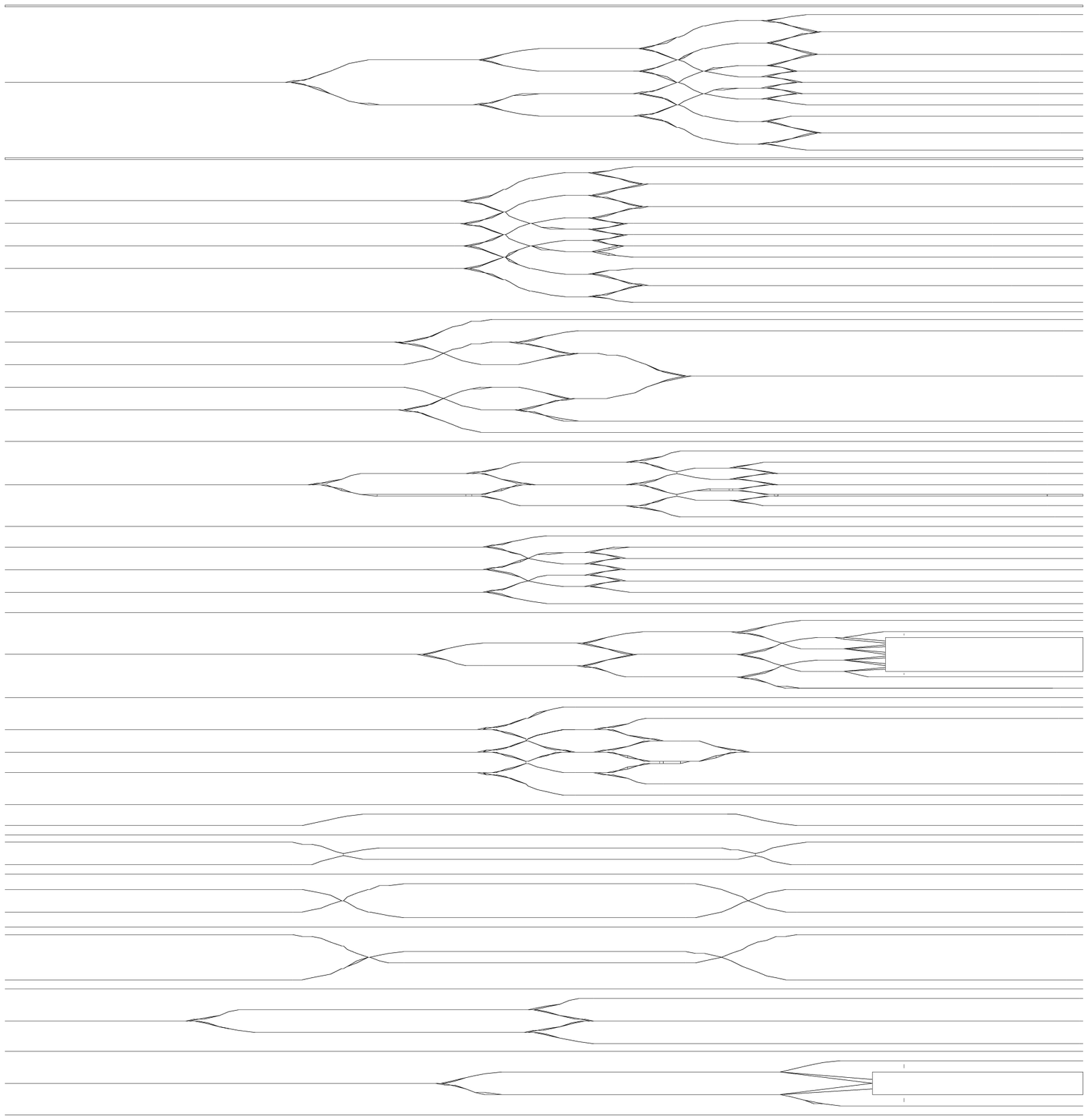}
\hspace*{1cm}\includegraphics[width=.15\textwidth]{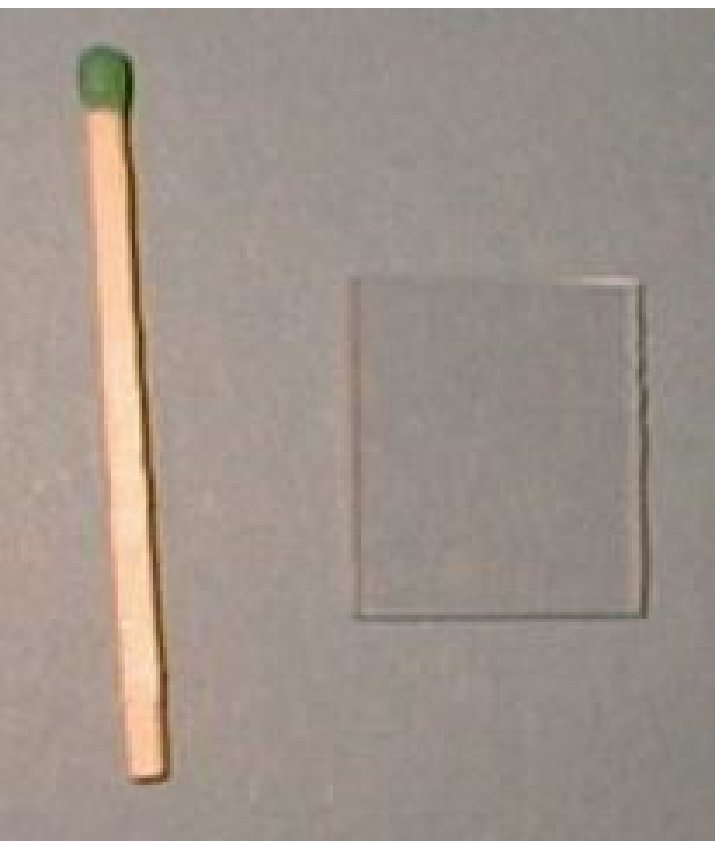}
\caption[]{Mask produced with LEMO facilities containing tests 
components for telescope beam combination (2, 3 and 4 telescope) 
either in coaxial and multi-axial mode (left). Half part of the chip 
(right), contains all the components.}
\label{LEMO-mask}
\end{figure}

The LEMO components allowed an analysis of several beam combinations
modes: Y junctions, multi-axial, MMI. 2, 3 and 4 telescopes beam
combiners were produced.  An example of mask can be seen in figure
\ref{LEMO-mask}.  IO technology allows cut the chip to at different
input guide position in order to test all the beam combiners
integrated stages. For a first series of experiments all the beam
combiners inputs were linked together thanks to suitable Y-junctions
\cite{Hag2}.  In this case interference signal depends only on
internal component behaviors independently of the feeding optics
resulting in a characterization of IO specific properties. The obtained
results show the low influence to external constraints on the
component, and the good symmetry of the optical path within the chip
\cite{Hag2}.

\subsection{Validation of the concept on the sky}

Observations using 2 telescope components have been successfully
performed using IOTA facilities at Mount Hopkins, Arizona 
\cite{Ber01}.  Further
observations using 3-telescope facilities are foreseen in a second
step to validate 3-input components operation.  Possible installation
at Chara is currently under consideration.

\section{Perspectives}
The proposed technology \cite{Ker2} shows many advantages for
instrumentation design applied to astronomical
interferometry. Furthermore it gives a unique solution to
the problem beam combination of an array with large number of
telescopes.  The full instrument concept can be adapted from the
planar optics structure and properties.

\subsection{Planar optics advantages}
Planar optics was proposed in the context of single mode 
interferometry. It offers solutions to outline limitations of the fibers. 
Fiber optics introduces decisive inputs with a reduction of the 
number of degrees of freedom of the instrumental arrangement 
and the modal filtering. Planar optics 
introduces additional arguments. Its application is limited to the 
combiner instrument itself and is not suitable for beam 
transportation and large optical path modulation. We can summarize 
its main advantages:
\begin{itemize}
    \item Compactness of the whole instrument (typically 40 mm x 5 mm).
    \item Low sensitivity to external constraints
    \item Implementation in a cryostat
    \item Extremely high stability
    \item No tuning or adjustment requirement but the signal injections in the component, 
    while all the instrument is embedded in a single chip
    \item Combiner alignment difficulty reported on the component mask design
    \item Reduced complications while increasing the number of telescopes, 
    all difficulty is reported on the mask design \cite{Ber2}
    \item Intrinsic polarization capabilities
    \item The major cost driver is reported on the mask design 
    and optimization phase. Existing component duplication may be realized at low cost.
\end{itemize}

\subsection{Application for aperture synthesis}
Extrapolation of the tested design to larger number of telescopes is
investigated \cite{Ber2}. All elements exist to propose 8-telescope 
combiner in an optimized design, as an interesting 
concept for the whole VLTI coherencing mode. It offers a unique 
imaging capability for large interferometric arrays.
This important issue imposes accurate phase stability inside the components 
who can be provided by IO.

\subsection{Fringe sensor}
For more than 3 telescope operation, mainly for imagery, 
a fringe tracker is mandatory for each baseline.
IO provides a compact solution, for instance with all the 
component outputs corresponding to the baselines, imaged on the same detector 
array, in a single cryostat leading to a significant system 
simplification.
\subsection{Fully integrated instrument}
The achievable compactness opens attractive 
solutions for fully cooled instrument. In most of the case a chilled 
detector is required, that needs to be installed inside a cryostat. 
Low temperature of the environment is required to improve the detector efficiency 
and to avoid pollution by background emission. An integrated instrument 
allows to replace the cryostat 
window with a fiber feed through.

The instrument is also confined in a protected volume, and then can be
locked in a tuned position. In this case the component outputs are imaged
directly on the detector array through relay optics. Even future optimized
design may not require any relay optics while gluing an array detector on
the substrate end \cite{Clau} or implementing a STJ device directly
on the substrate \cite{Feau}. We investigate technological points to be
solved for the installation of the whole instrument inside the camera
cryostat in front of the detector \cite{Rou2}. In our prototype a relay
optics is implemented in order to keep flexibility, for engineering tests.

\subsection{IO for larger wavelengths}
Extrapolation of the operating technology to larger wavelengths is an other
important issue.  The LEMO mask can be used directly to produce all
available components for K band operation. Material transmission is
compatible with our requirement up to $2.5\mu {\rm m}$ \cite{Scha}. The
development requires a tuning of the ion exchange parameters.
Extrapolation of the operating wavelength to the thermal IR, ($ > 2.5 \mu
{\rm m}$ ) is a more critical issue. A current analysis will identify
materials with sufficient transmission. At the present time no single mode
fiber optics for $10 \mu {\rm m}$ are available in catalogs, and even
laboratory components transmission imposes length shorter than a few
millimeters. Investigations are in progress to produce planar guides for
the N band ($10 \mu {\rm m} $) \cite{Laur}.

Thermal IR instrument, as MIDI for the VLTI or Darwin / IRSI (ESA space
interferometer), may include planar optics components as modal filters. The
instrument thermal constraints can be reduced thanks to an implementation
of the component inside a cryostat close to the detection head, and by
reducing optical interfaces between subsystem optical components. IO
solutions for thermal IR could allow a reduction of the modulation effects due to
system operation (background subtraction, OPD modulation). The
availability of small range
delay lines using planar optics (up to tens of $\mu {\rm m}$) for fringe
trackers could reduce thermal modulation of the environmental
background.  Furthermore instrumental thermal emission can be fully
controlled by design for the guided part.

\subsection{Space based applications}
IO is a very attractive solution for a space-based technological testbed.
The IO techniques allows, accurate, robust and extremely light
concept. Such concept is compatible with a prototype dedicated to principle
demonstrations for complex mission as Darwin or other foreseen preparation
mission.

\section{Acknowledgments}

The authors are grateful to Pierre Benech and Isabelle Schanen for 
their strong collaboration to this work and P. Pouteau, P. Mottier 
and M. Severi (CEA/LETI - Grenoble) for their contribution in LETI 
components realization, F. Reynaud for fruitful 
discussions, and to E. Le Coarer and P. Feautrier for the idea of 
combining IO and STJ. These works have partially been
funded by PNHRA/INSU, CNES, CNRS / Ultimatech and DGA/DRET.
%


\end{document}